\documentstyle[preprint,aps]{revtex}
\begin{document}

\draft
\title
{\bf Stark effect on the exciton complexes of individual quantum
dots}

\author{ David M.-T. Kuo$^{1}$ and Yia-Chung Chang$^{2}$}
\address{$^{1}$Department of Electrical Engineering, National Central University,\\
Chung-Li, Taiwan 320, Republic of China}


\address
{$^{2}$Department of Physics and Materials Research Laboratory\\
University of Illinois at Urbana-Champaign, Urbana, Illinois 61801}

\date{\today}
\maketitle

\begin{abstract}
The emission spectrum of exciton complexes formed in individual
self-assembled quantum dots (QDs) embedded into a p-n junction is
theoretically studied using an effective mass model. We calculate
the particle Coulomb interactions, eletron-hole overlaps and
transition energies of exciton complexes for the different
strength and direction of electric field. Both redshifts and
blueshifts are observed in exciton, trions and biexciton. The
Stark effect may be applied to manipulate the spontaneous emission
rate of individual QDs embedded in microcavities.
\end{abstract}

\newpage
\section{Introduction}
Recently, many efforts have devoted into the studies of
single-photon sources (SPS) composed of quantum dots (QDs), which
provide the potential applications of quantum cryptography and
quantum computing$^{1-3}$. For small size QDs, the strong three
dimensional confinement effect of QDs causes the large energy
level separation. Consequently, we can manipulate electrons and
holes to occupy the lowest energy level of QDs and operate emitted
photons from exciton, trion or biexciton formed in QDs. The
antibunching feature of SPS was demonstrated in ref[1,2] where
electros and holes of the QD are excited using optical pumping,
only a handful of studies employed the electrical pumping to
demonstrate the antibunching behavior of SPS.

From the practical point of view, the electrically driven SPS has
been suggested by Imamoglu and Yamamoto considering individual QDs
embedded into a semiconductor p-n junction$^{4}$. To fabricate a
single QD of nanometer size at specific location is one of
challenged techniques in the realization of electrically driven
SPS. Self-assembled quantum dots (SAQDs) combined with selective
formation method may solve such difficulty$^{5}$. Apart from the
electrically driven, it is very important to manipulate the the
spontaneous emission rate of SPS in the applications of quantum
cryptography. According to the Purcell effect$^{6-9}$, the
enhancement of the spontaneous emission rate due to cavity
($1/\gamma_{cav}$)with respect to free-space is
\begin{equation}
\frac{1}{\gamma_{cav}}=
\frac{1}{\gamma_{free}}\frac{3Q(\lambda_c/n)^3}{4\pi^2 V_{eff}}
\frac{\Delta\omega^2_c}{ 4(\omega-\omega_c)^2+\Delta \omega^2_c}
\frac{|\textbf{E}({\bf r}_e)|^2}{|E_{max}|^2} \eta^2 ,
\end{equation}
where $\eta=\textbf{d}
.\textbf{E}(\textbf{r})/|\textbf{d}||E(\textbf{r})|$ describes the
orientation matching of diploe $\textbf{d}$ and  electric field of
emission light $\textbf{E}$, and where the factor 3 from a 1/3
averaging factor accounting for the random polarization of
free-space modes with respect to the dipole . The quality factor
of cavity is given by $Q=\omega_c/\Delta\omega_c$, where
$\omega_c$ and $\Delta \omega_c$ denote, respectively,  the
angular frequency and linewidth of a cavity supporting a
single-mode. Other notations $V_{eff}$, and $n$ are the effective
volume of cavity and the refractive index at the field maximum.
Although a number of experiments have already demonstrated the
potential of QD-based solid-state cavity quantum electrodynamics
(QED) in applications such as single-photon sources$^{1-2,10-11}$.
Nevertheless, because it is very difficult to pre-determine the
exact resonance energy and location of an optically or
electrically active QD, all of the prior QD-based cavity QED
experiments relied on a random spectral and spatial overlap
between QDs and cavity modes. Most recently, in spite of  Badolado
et al have proposed a method to solve above difficulty$^{12}$. The
studies of optimization of Purcell effect for electrically driven
SPS are still lack.

Compared with optical pumping, the electrically driven SPS provide
a tunable emission spectrum of exciton complexes arising from the
Stark effect. Such extra degree of freedom (tunable wavelengthes)
provides a possibility to reach $\omega=\omega_c$ of eq.
(1)condition. Although the Stark effect on exciton peaks formed in
QDs was well investigated experimentally and
theoretically$^{13-19}$, the Stark effect on the trion and
biexciton peaks is still not very clear. Thompson et al.
demonstrated that the single photon generated by the biexciton
state is significantly less in emission time than the single
exciton state$^{20}$. From this point of view, the biexciton state
is favored for the application of single photon generation.
Therefore, it is worth studying the physical parameters of
biextion state.

The modest purpose of this article is to theoretically study the
Stark effect on the transition energies of exciton complexes.
Besides, we will also investigate the electric field effect on the
particle Coulomb interactions and oscillator strengthes of QDs.
The formal determines the charging energies for electrons and
holes, which are crucial parameters for the electrically driven
SPS. The latter influences the spontaneous emission rate of QDs in
free space.

\section{Formalism}
Because a number of experiments used InAs/GaAs SAQD system to
fabricate SPS, we consider such SAQDs embedded into a p-n junction
shown in Fig. 1. To understand the emission spectrum of an
InAs/GaAs SAQDs, we need to calculate the electronic structures
for various electron-hole complexes, including exciton ($X$),
biexciton ($X^2$), negative trion ($X^-$), and positive trion
($X^+$), which can occur in this system. Due to the large
strain-induced splitting between heavy-hole and light-hole band
for InAs on GaAs, we only have to consider the heavy hole band
(with $J_z=\pm 3/2$) and ignore its coupling with light-hole band
caused by the QD potential. Thus, we can treat the heavy hole as a
spin-1/2 particle with $\sigma=\uparrow, \downarrow$ representing
$J_z=\pm 3/2$.  Meanwhile, we adopt the Hartree-Fock (HF)
approximation to describe the many-particle systems for trions and
biexciton. This is a reasonable approximation for many particles
confined in a small QD, because the particle correlation effect is
greatly suppressed by the lack of available low-energy
excitations, which are coupled to the HF ground state via the
configuration interaction. In two-electron atomic systems such as
H$^-$, He and Li$^+$, it is well known that the correlation energy
for all these systems is around 0.11 Ry$^{21}$. A major part of it
can be accounted for within the unrestricted Hartee-Fock (UHF)
approximation by the so called ``in-and-out correlation'', which
arises because the two electrons can occupy two orbitals with very
different radii. The remainder of the correlation is due to the
``angular correlation'', which may be described by the coupling of
the HF ground state to the two-particle excited states with
non-zero single-particle orbital angular momentum (while keeping
the total orbital angular momentum to be zero)$^{22}$. Both the
``in-and-out correlation'' and the ``angular correlation'' are
significantly suppressed in the QD system with strong particle
confinement.

The strengths of $U_e$, $U_h$, and $U_{eh}$ in excitons, trions,
or biexcitons are determined via the self-consistent HF
calculation within a simple but realistic effective-mass model. We
consider an InAs/GaAs self-assembled QD (SAQD) with conical shape.
The QD resides on a monolayer of InAs wetting layer and the whole
system is embedded in a slab of GaAs with finite width. The slab
is then placed in contact with heavily doped p-type and n-type
GaAs to form a p-i-n structure for single photon generation.
Within the effective-mass model, the electron and hole in the
exciton ($X$), trions ($X^-$ or $X^+$), or biexciton ($X^2$) in
the QD are described by the coupled equations

\begin{eqnarray} && [-\nabla \frac {\hbar^2} {2m_e^*(\rho,z)} \nabla +
V^e_{QD}(\rho,z) - eFz+ V_{sc}(\rho,z)] \nonumber \\
&& \psi_e(\rho,\phi,z) = (E_e-E_c) \psi_e(\rho,\phi,z),
\label{HFe}
\end{eqnarray} \begin{eqnarray} && [-\nabla \frac {\hbar^2} {2m_h^*(\rho,z)}
\nabla +
V^h_{QD}(\rho,z) + eFz-V_{sc}(\rho,z)]\nonumber \\
&& \psi_h(\rho,\phi,z) = E_h \psi_h (\rho,\phi,z), \label{HFh}
\end{eqnarray} and
\begin{equation} V_{sc}({\bf
r})= \int d{\bf r}' \frac {e^2 [n_e({\bf r'})-n_h({\bf r}')]}
{\epsilon_0 |{\bf r}'-{\bf r}|}, \end{equation} where $E_c$
denotes the energy of the GaAs conduction band minimum, which is
$1.518 eV$ above the GaAs valence band maximum (defined as the
energy zero) , ${m_e^*(\rho,z)}$ (a scalar) denotes the
position-dependent electron effective mass, which takes on values
of $m_{eG}^* = 0.067 m_e$ for GaAs and $m_{eI}^* = 0.04
m_e$$^{23}$ for InAs. ${m_h^*(\rho,z)}$ denotes the
position-dependent effective mass tensor for the hole. Due to the
strong spin-orbit coupling and the large strain-induced splitting
between heavy-hole and light-hole band, we can neglect the
coupling of the heavy-hole band with the split-off band and
light-hole band. Consequently, it is a fairly good approximation
to describe ${m_h^*(\rho,z)}$ in InAs/GaAs QD as a diagonal tensor
with the $x$ and $y$ components given by
${m^*_t}^{-1}=(\gamma_1+\gamma_2)/m_e$ and the $z$ component given
by ${m^*_l}^{-1}=(\gamma_1-2\gamma_2)/m_e$. $\gamma_1$ and
$\gamma_2$ are the Luttinger parameters. Their values for InAs and
GaAs are taken from Ref. [24]. $V^e_{QD}(\rho,z)$
($V^h_{QD}(\rho,z)$) is approximated by a constant attractive
potential in the InAs region with value determined by the
conduction-band (valence-band) offset and the deformation
potential shift caused by the biaxial strain in the QD. These
values have been determined by comparison with results obtained
from a microscopic model calculation$^{25}$ and we have $V^e_{QD}
\sim -0.42 eV$ and $V^h_{QD} \sim -0.3 eV$. The $eFz$ term in
Eqs.~(\ref{HFe}) and (\ref{HFh}) arises from the applied voltage,
where $F$ denotes the strength of the electric field. $V_{sc}({\bf
r})$ denotes the self-consistent potential caused by the
electrostatic interaction with the charge densities [$n_e({\bf
r})$ and/or  $n_h({\bf r})$] associated with the other particles
in the system. $\epsilon_0$ is the static dielectric constant of
InAs. The image force is ignored here due to the small difference
in dielectric constant between InAs and GaAs.

Eqs.~(\ref{HFe}) and (\ref{HFh}) are solved self-consistently via
the Ritz variational method by expanding the ground-state wave
functions in terms of a set of nearly complete basis functions,
which are chosen to be products of Bessel functions (with axial
symmetry) and sine waves
\begin{equation}
\psi_{nlm}(\rho,\phi,z)= J_0(\alpha_n\rho) \sin (k_m
(z+\frac{L}{2})),
\end{equation}
where $k_m = m\pi/L$, m=1,2,3... $L$ is taken to be 300\AA. $J_0$
is the Bessel function of zero-th order and $\alpha_nR$ is the
$n$-th zero of $J_0(x)$ with $R$ taken to be 400\AA, which is
large enough to give  convergent numerical results. Forty sine
functions multiplied by fifteen Bessel functions are used to
diagonalize the Hartree-Fock Hamiltonian.

The ground state energy of  system $i \; (i=X,X^2,X^-,X^+)$is
given by
\begin{eqnarray} E_i=N^i_e E^i_e + N^i_h E^i_h + N^i_eN^i_h U^i_{eh}\nonumber
\\-N^i_eN^i_e U^i_e -N^i_hN^i_h U^i_h ,
\end{eqnarray} where $E^i_e$ and $E^i_h$ denote the ground state HF
single-particle energy for electron and hole obtained from solving
Eqs.~(\ref{HFe}) and (\ref{HFh}), respectively.  $N^i_e$ and
$N^i_h$ denote the electron and hole particle number in system
$i$. $U^{i}_e$, $U^{i}_h$ and $U^{i}_{eh}$ are the
electron-electron interaction, hole-hole interaction and
electron-hole interaction, respectively. The transition energies
for the electron-hole recombination in $X$, $X^2$, $X^-$, and
$X^+$ are give by $E_X$, $E_{X^2}-E_X$, $E_{X-}-E^0_e$, and
$E_{X+}-E^0_h$, respectively. Here $E^0_e$ and $E^0_h$ are the
free-particle ground state eigen-values for Eqs.~(\ref{HFe}) and
(\ref{HFh}) with $V_{sc}$ set to zero.

\section{Results}
For electrically driven SPS, electrons and holes are injected from
the electrodes. To create biexciton state, the applied voltage
needs to overcome the charging energies of electrons and holes,
which are determined by $U_e$ and $U_h$. The calculation of
$U_{e}$, $U_{eh}$, and $U_{h}$ is shown in Fig. 2, where the solid
curves denote $X^2$, the dotted curves denote $X^-$ , and the
dashed curves denote $X^+$ in conical SAQDs with base radius $R_0$
varying between 60 \AA \, and 110 \AA. Here, the QD height $h$ is
varied between 15 \AA \, and 65 \AA, while keeping the ratio
$(h-15\AA)/(R_0-60\AA)=1 $. The electron-hole Coulomb energy
$U_{eh}$ in the exciton $X$ is almost the same as that in $X^2$
and is omitted in this plot. The strengths of Coulomb interactions
are, in general, inversely proportional to the QD size, since the
charge densities in smaller QDs are more localized. However as the
QD size decreases below a threshold value (around $R_0$=70 \AA),
the Coulomb interactions becomes reduced as a result of the leak
out of charge density for small QDs, which tend to be more serious
for the electron than for the hole due to the smaller effective
mass for the electron. This manifests the effects of finite
potential barrier of the QD$^{26}$. This effect will not appear if
one adopts an infinite barrier approximation as in some previous
studies of excitons in QDs$^{27-29}$. In the large dot limit
$U_{e}$, $U_{eh}$ and $U_{h}$ all approach the same value,
indicating similar degree of localization for the electron and
hole in the large dot. Note that the magnitude of charging
energies for electrons or holes in the same order of thermal
energy of room temperature $k_B T$, where $k_B$ is a Boltzmann
constant. This indicates that the operation temperature of system
should be much lower than room temperature.

Owing to the applied bias crossing the QD, the electric field
effect is not negligible in the variation of particle
interactions. We adopt the size of QD with radius 7.5 nm and
height 3 nm to study the electric field effect on the particle
Coulomb interactions. We assume that the z axis is directed from
the base to the apex of the dot. Fig. 3 shows the Coulomb
interaction strengths as functions of electric field for different
exciton complexes. We see that $U_{h}$, $U_{eh}$ and $U_e$ display
asymmetric behavior of electric field as a result of geometer of
the dots. Increasing electric field, the deduction of $U_{eh}$
indicates that the electron-hole separation increases. However, we
note that $U_h$ increases in the positive direction of electric
field since the wave functions of holes become more localize. Even
though the enhancement of $U_e$ is observed in the negative
direction of electric field, it only exists at very small electric
field region. When the electric field is larger than a threshold
value, the wave functions of electrons become delocalize and leak
out the quantum dot. Consequently, electron-electron Coulomb
interactions becomes weak. As mentioned, $U_e$ and $U_h$ denote
the charging energies of QD for electrons and holes, respectively.
Therefore, the constant interaction model used extensively in the
Anderson model is valid only for small electric field case$^{30}$,
otherwise we should take into account bias-dependent Coulomb
interactions.

Based on eq. (1), the larger $1/\gamma_{free}$, the better for
$1/\gamma_{cav}$. Hence we also attempt to investigate the
oscillator strength of QDs, which is proportional to the
electron-hole overlap squared. Fig. 4 shows the squared
electron-hole overlap function $A(F=0)$ for $X,X^2,X^-$, and $X^+$
as functions of the QD radius, $R_0$ (with the height $h$ varying
in the same way as in Fig. 2). For all four complexes, the
electron-hole overlap increases with the QD size, indicating that
the electron and hole wave functions approach each other as they
become fully confined in the QD. The electron-hole overlaps in
exiton and biexciton are very similar since they are both charge
neutral. Meanwhile, the $A(F=0)$ of biexciton is slightly above
that of the exciton. For small size QDs, the electron-hole overlap
is the smallest in negative trion ($X^-$), and the largest in
positive trion ($X^+$). This is due to the fact that the Coulomb
repulsion between the two electrons in $X^-$ causes the electron
wave function to be more delocalized than in the charge-neutral
complexes ($X$ and $X^2$), thus reducing the electron-hole
overlap. However, the same effect in $X^+$ causes the hole wave
function to be more delocalized, which becomes closer to the
electron wave function, and therefore enhances the electron-hole
overlap.  The behavior is reversed for large QDs with a cross-over
occurring at $R_0 \sim 95$ \AA.

In fig. 4 we did not include the electric field effect for
electron-hole overlap function $A(F)$. Using the size of QD
considered in Fig. 3, we show $A(F)$ as functions of electric
field for four complexes in Fig. 5. $A(F)$ of each complex is
enhanced at small field region for $F>0$, whereas the decline of
$A(F)$ appears when the electric field is larger than a threshold
field $F_s$, which depends on exciton complexes. For example,
$F_s$ is about $45~kV/cm$ for $X^{+}$, but $30 ~kV/cm$ for $X$. On
the other hand, $A(F)$ declines as the electric field is increased
from 0 to $120 kV/cm$ for negative applied field direction. We
observe that the variation of $A(F)$ is the largest in negative
trion $X^{-}$ since the electron wave function is more delocalized
for $X^{-}$ than other complexes. Even though the results of Fig.
5 shows that $A(F)$ of positive trion and biexciton states is
larger than that of exciton at a given electric field. However,
the creation of biexciton (trion) state requires the applied field
higher than that of exciton for the overcome of charing energies,
so that it is possible to observe that the spontaneous emission
rate of biexciton or trion is smaller that of exciton in this
electrically driven SPS.

Finally, we attempt to study the transition energies for exciton
complexes without and with electric field. Fig. 6 shows the
transition energy for the electron-hole recombination in a
biexciton, positive trion and negative trion relative to the
exciton transition energy ($E_X$) as functions of the QD radius,
$R_0$ (with the height $h$ varying in the same way as in Fig. 2).
The exciton transition energy ($E_X$) as a function of $R_0$ is
also displayed (dashed curve) with the scale indicated on the
right side of Fig. 6. The biexciton transition energy is
consistently above the exciton transition energy ($E_X$) for the
range of $R_0$ considered here, which is still considerably
smaller than the free exciton Bohr radius in InAs ($\sim 300$\AA).
For very large QDs (with $R_0 > 300$\AA), the correlation effect
will become important, and the biexciton transition energy can
become lower than $E_X$.  The Positive (negative) trion appears at
significantly higher (lower) energy than the exciton for small
QDs, but approaches the biexciton quickly as the QD size
increases. This behavior can be understood by examining the
difference in $U_h$ ($U_e$) and $U_{eh}$ as shown in Fig. 2. The
biexciton peak displaying a blue shift with respect to the exciton
peak (showing an antibinding biexciton) is also consistent with
the observation reported in ref.[3,20]. Recently, studies of the
binding and antibinding of biexcitons were reported in Ref. 31.
Our calculation given by Eqs.~(2) and (3) provides only the
antibinding feature of biexciton. In Ref. 31 it is pointed out
that the biexciton complex changes from antibinding to binding as
the QD size increases. For QDs with dimension larger than the
exciton Bohr radius, the correlation energy becomes significant.
In this study, we have not taken into account the correlation
energy. This is justified as long as we restrict ourselves to
small QDs.

According to eq. (1), it is one of challenged techniques to obtain
$\omega=\omega_c$. If the frequencies of emitted photons
contributed from exciton complexes can be highly tuned, the
possibility of optimizing spontaneous emission rate will be
enhanced. Fig. 7 (a) and (b) shows Stark shifts of exciton
complexes with the strength and direction of the electric field
for different quantum dot size. In diagram (a) the size of QD
corresponds to that of Fig. 3. In diagram (b) we adopt the size of
QD with the radius $R_0=8.5 nm$ and height $h=3 nm$. The exciton
complexes display asymmetric shift around zero field and red shift
in the negative field direction. We see that in the positive field
direction the exciton complexes display the blue shifts at small
field region. The blue shift of positive trion is relatively large
if compare with that of other exciton complexes. For charge
neutral exciton $X$ and $X^2$, their responses to Stark effect are
almost the same. The comparison of diagram (a) and (b) shows that
the tunable range of emitted photon frequencies depends on the
size of quantum dot. In our case, Stark shifts of exciton
complexes are near $2.8~meV$ and $3.5~meV $ for the QD of the
radius $R_0=7.5 nm$ and height $h=3~ nm$ and the QD of the radius
$R_0=8.5 nm$ and height $h=4~ nm$, respectively. This remarkable
frequency shifts arising from the Stark effect lead the SPS with
electric pumping to readily reach the resonant condition of
$\omega=\omega_c$.

\section{Conclusion}
In this article we have calculated the following three physical
parameters of exciton complexes formed in the ground state of QD
embedded into a p-n junction: particle Coulomb interactions,
electron-hole overlaps and transition energies. In the typical
volume of grown SAQDs, we found that the magnitude of
electron-electron Coulomb interactions as well as hole-hole
Coulomb interactions is just near the thermal energy of room
temperature. Therefore, it is difficult for the InAs/GaAs system
to provide single-photon sources at room temperature. We note that
the suppression of electron-hole overlaps always exist for
electrically driven SPS. Although the decaying of biexciton to
exciton can generate single photons, we need to apply relatively
higher voltage to overcome the charging energies for electrons and
holes to create the biexciton state. Such higher voltage is
possible to suppress the spontaneous emission rate of the
biexciton state and make it smaller than that of exciton at lower
voltage. Consequently, the exciton state is preferable for
single-photon generation.

Owing to the asymmetric shape of QDs, the direction-dependent
Stark effects are observed in the measurable transition energies
of exciton complexes. We note that the Stark shifts of complexes
are smaller in the positive direction field than negative
direction field. Besides, the Stark shifts also depend on the size
of QDs. In particular, the largest blueshifts are occurred in the
positive trion state, on the other hand, the largest redshifts are
observed in the exciton state. The Stark shifts of exciton
complexes may be applied to manipulate the resonant condition of
eq. (1) and to control the spontaneous emission rate of individual
QDs embedded in a cavity.

\mbox{}

{\bf Acknowledgments}

This work was supported by National Science Council of Republic of
China Contract No. NSC 92-2112-M-008-053

\mbox{}

\newpage

{\bf Figure Captions}

Fig. 1: Energy-band structure of n-i-p junction for a cross
section through the quantum dot along z axis. The quantum dot is a
small conical structure located in the i layer between n and p
semiconductors. Electrons (black) and holes (white) to tunnel
cross the barriers with rate $\Gamma_e$ and $\Gamma_h$,
respectively. $R_{eh} $ is the electron-hole recombination rate.

Fig. 2: $U_{e}$, $U_{eh}$, and $U_{h}$ as a function of the QD
size for biexciton (solid), negative trion $X^{-}$ (dotted) and
positive trion $X^{+}$ (dashed). Note that the ratio
$(h-15\AA)/(R_0-60\AA)=1$ is used.

Fig. 3: Particle Coulomb interactions as a function of the
strength and direction of electric field for the QD with radius
$7.5~nm $ and height $3~nm$. Solid line denotes the biexciton
configuration, dashed line and dotted line denote, respectively,
positive trion and negative trion.

Fig. 4: Electron-hole overlap squared as a function of the QD size
for exciton $X$, biexciton $X^2$, negative trion $X^{-}$ and
positive trion $X^{+}$.

Fig. 5: Electron-hole overlap squared as a function of the
strength and direction of electric field for the QD with radius
$7.5~nm $ and height $3~nm$.

Fig. 6: Transition energy as a function of the QD size for exciton
$X$, biexciton $X^2$, negative trion $X^{-}$ and positive trion
$X^{+}$.

Fig. 7: Transition energy as a function of the strength and
direction of electric field for different QD size. Diagram (a) and
(b) denote, respectively, the QD of radius $7.5~nm $ and height $3
~nm$ and the QD of radius $8.5 ~nm $ and height $4~ nm$.

\end{document}